\documentclass[aps,prd,twocolumn,floatfix,nofootinbib,showpacs,superscriptaddress]{revtex4}
\usepackage{graphicx,amsmath,amssymb,bm}
\usepackage{color}

\definecolor{purple}{rgb}{0.5,0,0.5}
\definecolor{blue}{rgb}{0.0,0,0.9}
\usepackage[colorlinks=true, pdfstartview=FitV, citecolor= purple, linkcolor = blue,urlcolor=blue]{hyperref}
\newcommand{\feynslash}[1]{\slash\!\!\! #1}
\newcommand{\traceD}{\mathrm{tr}_\mathrm{CD}\!\ }
\newcommand{\gmor}{^\mathrm{ GMOR}}

\begin{document}

\title{Exciting flavored bound states}

\author{E.~Rojas}
\email{eduardo.rojas@cruzeirodosul.edu.br}
\affiliation{Laborat\'orio de F\'isica Te\'orica e Computacional, Universidade Cruzeiro do Sul, Rua Galv\~ao Bueno, 868, 01506-000 S\~ao Paulo, S\~ao Paulo, Brazil}
\author{B.~El-Bennich}
\email{bruno.bennich@cruzeirodosul.edu.br}
\affiliation{Laborat\'orio de F\'isica Te\'orica e Computacional, Universidade Cruzeiro do Sul, Rua Galv\~ao Bueno, 868, 01506-000 S\~ao Paulo, S\~ao Paulo, Brazil}
\author{J.~P.~B.~C.~de~Melo}
\email{joao.mello@cruzeirodosul.edu.br}
\affiliation{Laborat\'orio de F\'isica Te\'orica e Computacional, Universidade Cruzeiro do Sul, Rua Galv\~ao Bueno, 868, 01506-000 S\~ao Paulo, S\~ao Paulo, Brazil}

\begin{abstract}
We study ground and radial excitations of flavor singlet and flavored pseudoscalar mesons within the framework of the rainbow-ladder 
truncation using an infrared massive and finite interaction in agreement with recent results for the gluon-dressing function 
from lattice QCD and Dyson-Schwinger equations. Whereas the ground-state masses and decay constants of the light mesons as well as 
charmonia are well described, we confirm previous observations that this truncation is inadequate to provide realistic predictions for the 
spectrum of excited and exotic states. Moreover, we find a complex conjugate pair of eigenvalues for the excited $D_{(s)}$ mesons, 
which indicates a non-Hermiticity of the interaction kernel in the case of heavy-light systems and the present truncation. Nevertheless, 
limiting ourselves to the leading contributions of the Bethe-Salpeter amplitudes, we find a reasonable description of the charmed ground 
states and their respective decay constants.

\end{abstract}

\maketitle

\section{Introduction  \label{intro}}

Bound states are a fascinating subject as they teach us a great deal about the nature of interactions between constituents of a given
field theory. In the case of Quantum Chromodynamics (QCD), this fascination is augmented by the empirical fact that no asymptotically
free states,  i.e. colored quarks and gluons, are observed.  While studies of the analytic structure of colored Green functions are very 
instructive and were indeed crucial to get insight on their infrared behavior~\cite{Krein:1990sf,Bowman:2004jm,Bowman:2002bm,
Bowman:2005vx,Bhagwat:2003vw,Furui:2006ks,Cucchieri:2007zm,Cucchieri:2010xr,Bogolubsky:2009dc,Sternbeck:2012mf,Oliveira:2012eh,
Ayala:2012pb,Boucaud:2008ji,Boucaud:2011ug,Aguilar:2006gr,Aguilar:2008xm,Pennington:2011xs,Huber:2012kd,Blum:2014gna,
Fischer:2008uz,Strauss:2012dg,Dudal:2007cw,Dudal:2008sp,Dudal:2010tf,Cucchieri:2011ig,Oliveira:2010xc,Maas:2011se} and associated 
confinement mechanisms, they do not offer direct access to experimental  tests.  Nonetheless, their precise form may be studied indirectly 
via electromagnetic probes of the bound states formed by the theory's constituents~\cite{Holt:2010vj,Bashir:2012fs,Aznauryan:2012ba,Cloet:2013jya} 
and many dedicated experiments at existing and future accelerator facilities will serve exactly this purpose.

In QCD, the simplest possible bound state is given by the pion which plays a pivotal role in the understanding of the low-energy domain, 
being the lightest strongly bound antiquark-quark state  as well as the Goldstone bosons associated with chiral symmetry breaking. 
Model-independent properties of the Goldstone boson were derived long ago~\cite{Maris:1997hd} and express the intimate 
connection between the pion's Bethe-Salpeter amplitude (BSA) and the quark propagator in the chiral limit.

Heavy mesons, on the other hand, provide a formidable opportunity to study additional nonperturbative features of QCD and can be 
used to test simultaneously all manifestations of the Standard Model, namely the interplay between electroweak and strong interactions. 
In the infinite-heavy quark limit, the heavy-quark velocity becomes a conserved quantity and the momentum exchange with surrounding 
light degrees of freedom is predominantly soft. Since the heavy-quark spin decouples in this limit, light quarks are blind to it. In essence,
they do not experience any different interactions with a much heavier quark in a pseudoscalar or vector meson. In practice, however, 
heavy-flavor and heavy-spin breaking effects are important and while dynamical chiral symmetry breaking (DCSB) hardly plays a role in 
charmonium and bottonium states, it cannot  be ignored in heavy-light systems, such as in $D$ and $B$ mesons. A daunting challenge 
is presented by the disparate energy scales and asymmetric momentum distribution within these  flavor nonsinglet mesons, i.e.
the simultaneous treatment of heavy-quark symmetry breaking effects and DCSB, as the interactions of a heavy with a light 
quark are governed by the nonperturbative dynamics of the order $\Lambda_\mathrm{QCD}$. Incidentally, this is also 
the heavy meson's size scale.
 
Numerical solutions of the heavy meson's BSA with renormalization-group improved ladder truncation, following work on the kaon 
BSA~\cite{Maris:1997tm,Maris:1999nt}, proved to be unsuccessful: the truncations do not yield the Dirac equation when one of the 
quark masses is much larger. To circumvent these problems, the Dyson-Schwinger equation (DSE) for the heavy quark was solved 
for an infrared-suppressed gluon momentum as described in Ref.~\cite{Nguyen:2010yj}; no such infrared suppression was 
applied to the dressing of the light quark and the binding kernel. This approach reproduces well the masses for ground-state pseudoscalar 
and vector heavy-light mesons, but overestimates experimental weak decay constants by $\sim 20$\% in the case of the 
$D$ meson and $\sim 40$\% for the $B$ meson. Alternatively, in the heavy-quark propagation the mass function is approximated
by a momentum-independent constituent mass. This is justified for the $b$ quark and reasonable for the $c$ quark given the 
modest variation in its functional behavior over a wide momentum range and the constituent-quark mass is fitted to the lightest 
meson, the $D$ and $B$ mesons in the case of pseudoscalars~\cite{Nguyen:2010yj,Souchlas:2010zz}. The resulting BSA yields again 
ground-state pseudoscalar and vector meson masses which compare very well with experimental values but strongly underestimate 
the leptonic decay constants. The situation improves when one uses a complex conjugate pole representation (with three poles) 
instead of a heavy-quark constituent propagator to compute the normalization and weak decay constant,  though omitting a simultaneous 
application to the Bethe-Salpeter kernel seems inconsistent~\cite{Souchlas:2010zz}. Nevertheless, in the latter case the decay constants 
are overestimated by about 15\% for the $D$, $D_s$ and $B$ mesons compared with experimental averages.

Aside from an unfavorable comparison with experimental data, one may object that the constituent-quark approach neglects valuable 
information of the quark's dressed mass function, namely, its imaginary part, 
$\mathrm{Im}\, M(p^2) \sim 0$~\cite{ElBennich:2008qa,ElBennich:2009vx,ElBennich:2008xy,ElBennich:2012ij}.
Thus, in solving the Bethe-Salpeter equation~(BSE), one approximates the complex mass function by a flat surface on the complex plane. 
Our motivation is to verify whether a fully consistent, simultaneous  numerical solution of the heavy-quark DSE and heavy-light
pseudoscalar meson BSE can be obtained with a modern approach to the rainbow-ladder~(RL) truncation based on the interaction 
proposed by Qin {\em et al.\/}~\cite{Qin:2011dd}. This {\em Ansatz}  produces an infrared behavior of the interaction, commonly described 
by a ``dressing function'' $\mathcal{G}(k^2)$~\cite{Maris:1997hd}, congruent with the decoupling solution found in DSE 
and lattice studies of the gluon propagator~\cite{Cucchieri:2007zm,Cucchieri:2010xr,Bogolubsky:2009dc,Sternbeck:2012mf,
Oliveira:2012eh,Ayala:2012pb,Boucaud:2008ji,Boucaud:2011ug,Aguilar:2006gr,Aguilar:2008xm,Pennington:2011xs,Strauss:2012dg}. 
Indeed, the gluon propagator is found to be a bounded and regular function of spacelike momenta with a maximum 
value at $k^2 = 0$.  

Obviously, the cause for the difficulties encountered in solving the BSE for heavy-light mesons may not solely lie in the infrared
behavior of the interaction but also in the truncation employed. In particular, hadron properties are not much affected by the deep
infrared of the interaction; rather, they seem to be sensitive to the support and strength of the interaction at the scales of a few 
hundred MeV~\cite{Blank:2010pa,Bashir:2012fs}. As known from a long series of BSE studies, the RL truncation is 
successful for equal-mass mesons, for instance light $\bar qq$ mesons, such as the pion, kaon and  $\rho$~\cite{Maris:1997hd}, 
but also for $\bar QQ$ charmonia and bottonium~\cite{Maris:2005tt,Krassnigg:2004if,Bhagwat:2004hn,Blank:2011ha} where
subtle cancellations between contributions from the quark-gluon and antiquark-gluon vertices are in order.  Due to the very different 
momentum and energy-scale distributions, the dressing of the quark-gluon vertex becomes important for heavy-light systems. 
Progress to go beyond RL truncation has been made and is under way~\cite{Bender:1996bb,Eichmann:2008ae,Fischer:2008wy,
Fischer:2009jm,Qin:2013mta,Heupel:2014ina,Rojas:2013tza} and the consistent inclusion of a dressed quark-gluon vertex which 
satisfies the relevant Ward-Takahashi identities (WTI) can be realized by a general form of the BSE~\cite{Chang:2009zb}. 

Nonetheless, we here focus on the interaction proposed in Ref.~\cite{Qin:2011dd} within the RL approximation to describe the properties 
of the light, strange and charmed flavor singlet and nonsinglet pseudoscalar mesons, such as their mass spectrum, that of the first radial 
excitation and related weak decay constants. As we demonstrate, recent improved numerical techniques~\cite{Fischer:2005en,Krassnigg:2009gd} 
and a careful numerical treatment of the flavored DSE and BSE in conjunction with a particular infrared behavior of the interaction is able to 
satisfactorily reproduce the aforementioned observables. In particular, we compute the decay constants of the radial excitations of the kaon, 
$\eta_c$ and a fictitious $\bar ss$ state. We recall that this was not possible for the heavy-light mesons within the conventional RL framework of 
Ref.~\cite{Nguyen:2010yj}. In the present work, solutions for the $D$ and $D_s$ mesons are found, though the truncation leads to complex 
eigenvalues of their excited states if higher moments of the Chebyshev expansion of the BSA are included. We thus confirm that an a accurate 
and veracious  description of the heavy-light flavored mesons requires BSE solutions beyond the RL approximation.

\section{Pseudoscalar Bound States}

\subsection{Dyson-Schwinger equation}

We work in a continuum approach of QCD based on the RL truncation of the quark DSE and of the quark-antiquark kernel of the 
meson's BSE, which is the leading term in a systematic symmetry-preserving truncation scheme. The quark's gap equation is 
described by the following DSE,\footnote{We employ throughout a Euclidean metric in our notation: 
$\{\gamma_\mu,\gamma_\nu\} = 2\delta_{\mu\nu}$; $\gamma_\mu^\dagger = \gamma_\mu$; $\gamma_5= 
\gamma_4\gamma_1\gamma_2\gamma_3$, tr$[\gamma_4\gamma_\mu\gamma_\nu\gamma_\rho\gamma_\sigma]=-4\,
 \epsilon_{\mu\nu\rho\sigma}$; $\sigma_{\mu\nu}=(i/2)[\gamma_\mu,\gamma_\nu]$; 
$a \cdot b = \sum_{i=1}^4 a_i b_i$; and $P_\mu$ timelike $\Rightarrow$ $P^2<0$.}
\begin{equation}
S^{-1}(p)  =   \, Z_2 (i\, \feynslash  p + m^{\mathrm{bm}}) + \Sigma (p^2) \ ,
\label{DSEquark}
\end{equation}
where the dressed-quark self-energy contribution is ($q=k-p$),
\begin{equation}
\label{dressedquark}
 \Sigma (p^2) = Z_1\, g^2\!\! \int^\Lambda_k \!\!  D^{\mu\nu} (q) \frac{\lambda^a}{2} \gamma_\mu\, S(k) \,\Gamma^a_\nu (k,p) \ .
\end{equation}
The mnemonic shorthand $\int_k^\Lambda\equiv \int^\Lambda d^4k/(2\pi)^4$ represents a Poincar\'e-invariant regularization of the 
integral with the regularization mass scale, $\Lambda$, and where $Z_{1,2}(\mu,\Lambda )$ are the vertex and quark wave-function 
renormalization constants. The nonperturbative interactions contribute to the self-energy, $\Sigma (p^2)$, which corrects  the 
current-quark bare mass, $m^{\mathrm{bm}}(\Lambda)$. The integral is over the dressed gluon propagator, $D_{\mu\nu}(q)$, and
the dressed quark-gluon vertex, $\Gamma^a_\nu (k,p)$, and the color matrices $\lambda^a$ are in the fundamental representation
of SU(3). In Landau gauge the gluon propagator is purely transversal,
\begin{equation}
   D^{ab}_{\mu\nu} (q) =  \delta^{ab} \left( g_{\mu\nu} - \frac{k_\mu k_\nu}{q^2} \right) \frac{\Delta ( q^2) }{q^2} \ ,
\end{equation}   
where $\Delta ( k^2)$ is the gluon-dressing function. The quark-gluon vertex in this truncation is simply given by,
\begin{equation}
  \Gamma^a_\mu (k, p) =  \frac{\lambda^a}{2} \, \gamma_\mu  \  .
\end{equation}
Following Ref.~\cite{Qin:2011dd} and suppressing color indices, we write
\begin{equation}
  Z_1 g^2 \, D_{\mu\nu} (q) \,  \Gamma_\mu (k, p) \ \rightarrow \  Z_2^2\, \frac{\mathcal{G} (q^2)}{q^2} \, T_{\mu\nu} (q) \,  \gamma_\mu \ ,
\end{equation}
with the transverse projection operator $ T_{\mu\nu} (q) :=  g_{\mu\nu} -  q_\mu q_\nu/q^2 $ and the effective coupling is 
the sum of two terms:
\begin{equation}
\label{qinchang}
  \frac{ \mathcal{G} (q^2)}{q^2} =   \frac{8\pi^2}{\omega^4}  D  \exp \left ( -  \frac{q^2}{\omega^2} \right ) +
                  \frac{8\pi^2 \gamma_m\, \mathcal{F}(q^2) }{\ln \left [\tau + \left (1 + q^2/\Lambda_\mathrm{QCD}^2 \right )^2 \right ] } \ .
\end{equation}
In Eq.~(\ref{qinchang}), $\gamma_m = 12/(33 - 2N_f )$, $N_f = 4$, $\Lambda_\mathrm{QCD} = 0.234$~GeV;
$\tau = e^2 - 1$; and $\mathcal{F}(q^2) = [1 - \exp(-q^2/4m^2_t  ) ]/q^2$, $m_t = 0.5$~GeV. The parameters $\omega$ and $D$, 
respectively, control the width and strength of the interaction and appropriate values will be discussed in Sec.~\ref{discussion}. 
The second term is a bounded, monotonically decreasing and regular continuation of the perturbative-QCD (pQCD) running coupling for all 
spacelike values of $k^2$, whereas the first term is an {\em Ansatz\/} for the interaction at infrared momenta. It is crucial that the 
form of this  {\em Ansatz\/} provides enough strength to realize DCSB and/or confinement.  For momenta $k^2 \gtrsim 2$~GeV$^2$, 
the perturbative contributions markedly dominate the interaction. 

With this, the solutions for spacelike momenta, $p^2>0$, to the gap equation~(\ref{DSEquark}) for a given flavor, $f$, are generally,
\begin{equation}
   S_f(p) =  \left [ i\, \feynslash  p\ A_f(p^2) +\mathbb{I}_D\,  B_f(p^2) \right ]^{-1} \  ,
  \label{sigmaSV}
\end{equation}
where one imposes the renormalization condition,
\begin{equation}
\left. Z_f (p^2) = 1/A_f (p^2)  \right |_{p^2 = \mu^2} = 1 \ , 
\label{EQ:Amu_ren}
\end{equation}
at large spacelike $\mu^2\gg \Lambda_\mathrm{QCD}^2$. The mass function,  $M_f(p^2)=B_f(p^2\!, \mu^2)/A_f(p^2\!, \mu^2)$, is 
independent of the renormalization point $\mu$. In order to make quantitative matching with pQCD,  another renormalization condition,
\begin{equation}
\left.  S^{-1}_f(p) \right |_{p^2=\mu^2}  = \  i\  \feynslash  p \ + m_f(\mu )\, \mathbb{I}_D  \ ,
\label{massmu_ren}
\end{equation}
is imposed, where $m_f(\mu )$ is the renormalized running quark mass,
\begin{equation}
\label{mzeta} 
   Z_m^f  (\mu,\Lambda )\, m_f(\mu)  =  m_f^{\rm bm} (\Lambda) \  , 
\end{equation}
where $Z_m^f(\mu,\Lambda ) =  Z_4^f(\mu,\Lambda )/Z_2^f (\mu,\Lambda )$ is the flavor dependent mass-renormalization constant
and  $Z_4^f(\mu,\Lambda )$ is the renormalization constant associated with the Lagrangian's mass term. In particular, $m_f(\mu )$ 
is nothing else but the dressed-quark mass function evaluated at one particular deep spacelike point, $p^2=\mu^2$, namely:
\begin{equation}
  m_f(\mu)  = M_f(\mu )\,.
\end{equation}
The renormalization-group invariant current-quark mass can  be inferred via,
\begin{equation}
  \hat m_f =  \lim_{p^2 \to \infty} M_f (p^2) \left [ \frac{1}{2} \ln\left ( \frac{p^2}{\Lambda_\mathrm{QCD}^2}\right ) \right ]^{\gamma_m}  .
  \label{invariantmass}
\end{equation}

\subsection{Bethe-Salpeter equation}

The homogeneous  BSE for a $\bar qq$ bound state with relative momentum $p$ and total momentum $P$  can be written as,
\begin{equation}
\label{BSE} 
  \Gamma_{mn}^{fg} (p,P) = \int^\Lambda_k\! \!\!  \mathcal{K}^{kl}_{mn}(p,k,P)\left  [ S_f (k_+)\Gamma^{fg} (k,P) S_g (k_-) \right ]_{lk} ,
\end{equation}
where $m,n,k,l$ collect Dirac and color indices; $f,g$ are flavor indices; and $k_+ = k+\eta_+ P, k_- =k- \eta_- P; \eta_+ +\eta_- =1$.  
Since we work within the RL truncation, the BSE kernel  is given by
\begin{equation}
\label{BSEkernel} 
   \mathcal{K}^{kl}_{mn}(p,k,P) =  - \frac{Z_2^2\, \mathcal{G} (q^2 )}{q^2}
    \left (\frac{\lambda^a}{2}  \gamma_\mu \right )_{\!\!kn} \!\!\! T_{\mu\nu}(q) \,
     \left (\frac{\lambda^a}{2}  \gamma_\nu \right )_{\!\! ml } \! \!\! ,
\end{equation}
which satisfies the axial-vector WTI~\cite{Maris:1997hd} and therefore ensures a massless pion in the chiral limit. Eqs.~(\ref{BSE}) and 
(\ref{BSEkernel}) define an eigenvalue problem with physical solutions at the mass-shell points, $P^2 = -M^2$, where $M$ is the bound-state 
mass. 

In the following,  we consider the spectrum of flavor singlet and nonsinglet pseudoscalar mesons, which are exhibited as a pole 
contribution to the axial-vector and  pseudoscalar vertices (omitting regular terms in the neighborhood of the poles),
\begin{align}
 \Gamma_{5\mu}^{fg} (p,P)\big |_{P^2 +M_n^2 \simeq 0} & = \frac{f_{P_n}P_\mu}{P^2 +M_n^2}\,  \Gamma_{P_n}^{fg}  (p,P) \label{gamma5mu} \ ,  \\
  i \, \Gamma_{5}^{fg} (p,P)\big |_{P^2 +M_n^2 \simeq 0} & =  \frac{\rho_{P_n}}{P^2 +M_n^2}\,  \Gamma_{P_n}^{fg}  (p,P) \ ,
 \label{gamma5}
\end{align} 
where $\Gamma_{P_n}^{fg}  (p,P) \equiv \left [\Gamma_{mn}^{fg} (p,P)\right ]_{P_n}$ is the bound state's BSA. The principal quantum 
number is $n=0$ for the ground state and $n \geq 1$ for the radial excitations with bound-state mass,~$M_n$. In Eqs.~\eqref{gamma5mu} and
\eqref{gamma5}, the expressions for  $\rho_{P_n}$  and $f_{P_n}$ are defined in Eqs.~\eqref{rho} and \eqref{decay}, 
respectively.
The properties of the excited states are expected to be sensitive to details of the long-range component  of the interaction in Eq.~(\ref{qinchang}), 
which provides more support at large interquark separation than, e.g., the Maris-Tandy model~\cite{Maris:1999nt}. The present study thus allows 
for a test of the interaction of strange- and charm-flavored $0^-$ mesons and their excited states, in addition to those considered in
Ref.~\cite{Qin:2011dd}, via comparison with experimental values where available.

As mentioned in Sec.~\ref{intro}, it is most likely that beyond-RL contributions are important in heavy-light mesons. In fact, the RL truncation with 
the interaction in Eq.~(\ref{qinchang}) should work best for heavy-heavy ($\bar QQ$) systems. Consider for instance the Ball-Chiu {\em Ansatz\/} to the 
nonperturbative quark-gluon vertex, $\Gamma_\mu (k,p)$, which is fully determined with the knowledge of $A(p^2)$ and 
$B(p^2)$~\cite{Ball:1980ay,Ball:1980ax}:
for heavier flavors, $f=c, b$, the scalar function $B_c(p^2)$ and even more so $B_b(p^2)$ remain constant over a wide momentum 
range, while for the vector functions $A_{c,b}(k^2) \simeq A_{c,b}(p^2) \simeq  1$. The heavy-quark gluon vertex can therefore reasonably 
be  approximated by a bare vertex.

Thus, the RL approximation ought to work better for heavy mesons such as the charmonia and bottonium, which is indeed confirmed
in numerical studies~\cite{Bashir:2012fs,Krassnigg:2004if,Blank:2011ha}. The effects of DSCB and the importance of other tensor structures
in the quark-gluon vertex become increasingly more important for lighter quarks, where, for example, mass splittings between parity 
partners are only satisfactorily described by including the tensor structure which corresponds to the chromomagnetic moment~\cite{Chang:2011ei}. 
Nonetheless, a hallmark of the RL is the successful description of the pion, the key point being that the  axial-vector WTI be preserved and 
likewise the Goldstone-boson character of the pion.

\section{Numerical implementation}

\subsection{DSE on the complex plane \label{complexplane} }

We follow Refs.~\cite{Fischer:2005en,Krassnigg:2009gd,Fischer:2009jm} and choose the gluon momentum, $q$, to be real. This implies 
that when the external DSE momentum~(\ref{DSEquark}) is complex, so is the internal quark's momentum. It is the case in Eq.~(\ref{BSE}) 
which requires DSE solutions for $S(p_\pm)$ in the complex plane with,
\begin{equation}
   p_\pm^2 =  p^2 - \eta_\pm^2 M^2_n \pm  2i \sqrt{p^2}\eta_\pm M_n \ ,
   \label{externalmomenta}
\end{equation}
where $p$ is collinear with $P=({\vec 0}, iM_n)$ in the meson's rest frame and in the quark's DSE we employ $\eta_\pm =1/2$. 
As we add $P$ to the quark's external momentum, $p$, the internal quark momentum becomes $k_\pm = k\pm \eta_\pm P$, $k \in \mathbb{R}_+$, 
and it follows that $q=k_\pm-p_\pm=k-p$ is real. The internal quark's squared momentum is given by,
\begin{equation}
   k_\pm^2 =  k^2 - \eta_\pm^2 M^2_n \pm  2i \sqrt{k^2}\eta_\pm M_n z \ ,
 \label{parabola}
\end{equation}
with the angle, $-1\leq z \leq +1$, and is also bounded by a parabola whose vertex is $(0 , -\eta_\pm^2 M_n^2)$. Within the parabola,
the real integration variable, $k$, is limited by $\Lambda$ as follows.

We denote the regularization mass scale in Eq.~(\ref{BSE}) by $\Lambda_\mathrm{BSE}$ and relate it to the one in Eq.~\eqref{dressedquark} with,
\begin{equation}
  \Lambda_\mathrm{DSE}^2 = \Lambda_\mathrm{BSE}^2 +\tfrac{1}{4}\, M_f^2  = \Lambda_\mathrm{BSE}^2 + \eta_f^2 M_P^2 \ , 
\label{cutoff}
\end{equation}
for $f = u,d,s,c$. This procedure is best illustrated with an example,  e.g. for a meson with strangeness, $\bar u s$; we determine $\eta_s$ 
and $\eta_u$ pushing in each case $M_{f=u,s}$ to the limit at which we encounter poles while solving the DSE on the complex plane. 
From the latter equality in Eq.~(\ref{cutoff}), we obtain $\eta_s$ and $\eta_u$ and imposing $\eta_s+\eta_u =1$ we find,
\begin{equation}
\hspace*{-2mm}
  \tfrac{1}{2} M_s = \eta_s M_P \, ; \,   \tfrac{1}{2} M_u = \eta_u M_P \Rightarrow  \frac{M_s + M_u}{2} = M_P \, ,
  \label{maxmass}
\end{equation}
where $M_P$ is the ``maximal meson mass'' --- not to be confused with $M_n$ --- for which the DSE can be solved for each flavor and respective 
parameters, $\eta_s=M_s/(M_u+M_s)$ and  $\eta_u=M_u/(M_u+M_s)$. The latter are used as momentum partitioning parameters $\eta_\pm$ in the 
BSE. Of course, this is done for computational convenience and we have checked that our results are stable under variations of $\eta_\pm$, 
as expected from a Poincar\'e-invariant approach. In solving the BSE, Eq.~(\ref{maxmass}) imposes an upper boundary on the mass of the 
meson and its excited states: $M_n < M_P$. For flavor singlet equal-mass $\bar qq$ mesons we simply choose $\eta_f= \eta_+=\eta_-=1/2$.

The solution of Eq.~(\ref{DSEquark}) for a complex value of $p^2_\pm$ requires the simultaneous iteration on a complex $2d$ grid bounded
by the parabola of Eq.~(\ref{parabola}). Alternatively, we can exploit the information from the contour of the parabola via the Cauchy
theorem~\cite{Krassnigg:2009gd}. To proceed with the integration, we parametrize the contour (counterclockwise) in the upper complex plane, 
$\mathcal{C}_+ : p_+^2$, with $ p := z_+(t)= tp_{\text{min}}+(1-t)\Lambda_\mathrm{DSE}$; in the lower plane, $\mathcal{C}_- :  p_-^2$, with 
$p : = z_-(t)= t\Lambda_\mathrm{DSE} +(1-t)p_{\text{min}}$; and we close the parabola with the path $\mathcal{C}_y :  p_y^2 := \Lambda^2_\mathrm{DSE} - 
 y^2(t) +  2i\, y(t) \Lambda_\mathrm{DSE} $ with $y(t)=  M_n t  -\eta_-M_n $, $t \in [0,1]$.  
More precisely, we use a linear parametrization given by:
\begin{align}
  & dp_+^2  =  2\left ( z_+(t) +i\eta_+M_n \right )dz_+(t)\ ; \   0\leq z_+(t) < \Lambda_\mathrm{DSE}\ , \nonumber \\
  & dp_y^2 =  2\left (-y(t) +i \Lambda_\mathrm{DSE} \right ) dy(t) \, ; \, -\eta_- M_n\leq y(t)\leq \eta_+M_m, \nonumber \\
  &  dp_-^2  =  2\left ( z_+(t)  - i\eta_+M_n \right )dz_-(t)\ ; \   0\leq z_-(t) < \Lambda_\mathrm{DSE}\ . 
  \label{contour}
\end{align}
The DSE for the quark is solved for 128 complex momenta, $p_\pm$, on each contour section using 128 momenta, $k_\pm$, and 128 angles, $z$, in 
the interior of the parabola parametrized by Eq.~\eqref{contour}, where the values for $A(k_\pm^2)$ and $B(k_\pm^2)$ are obtained
via the Cauchy theorem and simultaneous iteration of their values on the contour. For numerical precision, the points on the contour 
are skewed towards the vertex of the parabola near $k^2_\pm \sim -\eta_\pm^2 M_n^2$. We use a quadrature method for the integral
evaluation and cross-check the results with the numerical integration library \texttt{CUBA}~\cite{Hahn:2004fe}. The BSE~\eqref{BSE} is solved
for 64 external momenta, $p$, 64 internal momenta, $k$, and 20 angles. 

The renormalization condition in Eqs.~(\ref{EQ:Amu_ren}) and (\ref{massmu_ren}) are imposed on the DSE solutions at the spacelike 
real-axis point, $\mu =19$~GeV, a value chosen to match the bulk of extant studies~\cite{Maris:2005tt,Qin:2011xq}.  The current-quark 
masses are fixed by requiring that the pion and kaon BSEs produce $m_\pi \approx 0.138$~GeV and $m_K \approx 0.493$~GeV. We thus 
use $m_{u,d}(\mu) = 3.4~\mathrm{MeV}$, $m_s(\mu) = 82~\mathrm{MeV}$~\cite{Qin:2011dd} and  $m_c(\mu) = 0.905$~GeV~\cite{Holl:2004fr}. 
However, in computing the values of $A(p^2)$ and $B(p^2)$ for the momenta on the parabola~\eqref{externalmomenta},
we are numerically limited by the appearance of conjugate poles and cannot use an arbitrarily large cutoff $\Lambda$ in Eq.~\eqref{DSEquark}. 
This implies that we impose the renormalization conditions, Eqs.~\eqref{EQ:Amu_ren} and \eqref{massmu_ren}, at a lower momentum value, 
$\xi < \mu$,  and thus obtain new renormalization constants, $Z_m^f$ and $Z_4^f$. The details of our renormalization procedure are discussed in
Appendix~\ref{renorm} and our complex solutions for $A(p^2)$ and $B(p^2)$ with the interaction in Eq.~\eqref{qinchang} are plotted for the 
$u$ quark in Fig.~3 of Ref.~\cite{El-Bennich:2013yna}, where the real parts of $A(p^2)$ and $B(p^2)$ on the parabola are described by equipotential 
lines with decreasing magnitude for increasing values of spacelike $\textrm{Re}~k^2 >0$. The imaginary components of $A(p^2)$ and $B(p^2)$
are negative valued and decreasing for $\textrm{Im} \ k^2 >0$ and positive and increasing for $\textrm{Im} \ k^2 < 0$, while for real momenta, 
$\textrm{Im}\ k^2 = 0$ and $\textrm{Re}\ k^2>0$, both functions are naturally real. 

Alternatively, we cross-checked our renormalization procedure by setting $A(\mu^2) = 1$ for $\mu = 2$~GeV and imposing in the ultraviolet 
the condition~\eqref{invariantmass} following Refs.~\cite{Qin:2011dd,Qin:2011xq}. We also reproduce our results using a complex pole representation
for the quarks with three poles~\cite{Chang:2013pq} to which we fit the DSE solutions on the real axis renormalized at 19~GeV. The numerical 
differences with the masses and decay constants obtained with the solutions of the DSE on the complex plane are negligible.

\subsection{Solving the BSE with Arnoldi factorization}

The general Poincar\'e-invariant form of the solutions of Eq.~(\ref{BSE}) for the  pseudoscalar channel $J^P= 0^-$ 
and the trajectory, $P^2= -M_n^2$, in a nonorthogonal base with respect to the Dirac trace, $\mathcal{A}^\alpha (p,P) = 
\gamma_5 \big \{ i\,\mathbb{I}_D$, $\feynslash P$, ${\feynslash p} (p\cdot  P)$, $\sigma_{\mu\nu}p_\mu P_\nu \big \}$, 
is given by,
\begin{align}    
  \Gamma_{P_n} (p, P)  =  \gamma_5\big [\, i\, \mathbb{I}_D  E_{P_n} (p,P) +  \feynslash P F_{P_n} (p,P) &  \nonumber  \\ 
                +\ \feynslash p (p\cdot P) G_{P_n} (p,P) + \sigma_{\mu\nu}p_\mu P_\nu\,   H_{P_n} (p, &  P) \big ]  \ , 
\label{diracbase}                            
\end{align}
where we have suppressed color, Dirac and flavor indices for the sake of visibility. The functions $\mathcal{F}_{P_n}^\alpha (p,P) =$ 
$\big \{ E_{P_n} (p,P), F_{P_n} (p,P), G_{P_n} (p,P), H_{P_n} (p,P)\big \}$  are Lorentz-invariant scalar amplitudes. We solve the 
BSE by projecting  on the functions $\mathcal{F}_{P_n}^\alpha (p,P)$ with the appropriate projectors, 
\begin{align}
    P^\alpha (p,P)  = \sum_{\beta=1}^4 P^{\alpha\beta}(p,P)\, \mathcal{A}^\beta (p,P)& \nonumber \\
   \tfrac{1}{4} \sum_{\beta=1}^4 P^{\alpha\beta}(p,P)\, \mathrm{Tr}_\mathrm{D}
    \left [ \mathcal{A}^\alpha (p,P)\mathcal{A}^\gamma (p,P)\right ]  & = \delta_{\alpha\gamma}\, ,
\end{align}
which allows for an extraction of $\mathcal{F}_{P_n}^\alpha (p,P)$  from the BSA,
\begin{equation} \hspace*{-1mm}
   \mathcal{F}_{P_n}^\alpha (p,P) =  \tfrac{1}{4}\, \traceD \big [ P^\alpha(p,P)\,\Gamma_P (p,P)  \big ]   \ .
\end{equation}
This leads to the eigenvalue equation ($\alpha,\beta =1,...,4$), 
\begin{equation}
  \lambda_n(P^2)\, \mathcal{F}_{P_n}^\alpha (p,P) =\int^\Lambda_k \!\! \mathcal{K}^{\alpha\beta} (p,k,P)\, \mathcal{F}_{P_n}^\beta (k,P) \ ,
 \label{eigenvalue}
\end{equation}
where  $\mathcal{K}^{\alpha\beta} (p,k,P)$ is obtained using Eq.~(\ref{BSEkernel}):
\begin{align}
  \mathcal{K}^{\alpha\beta} (p,k,P) & = -\frac{Z_2^2}{4} \frac{\mathcal{G} (q^2 )}{q^2}\, T_{\mu\nu}(q)\,  \traceD \Big [ P^\alpha (p,P) \nonumber \\
   &\times   \gamma_\mu\lambda^a  S (k_+)\,  \mathcal{A}^\beta (k,P)\, S (k_-)\,  \gamma_\nu\lambda^a  \Big ] \ .
\label{KernelAB}
\end{align}
The propagators and functions $\mathcal{F}_{P_n}^\alpha (p,P)$ in Eq.~(\ref{eigenvalue}) are expanded in Chebyshev polynomials of the 2nd kind, 
$U_m(z_k)$ and $U_m(z_p)$, with the angles, $z_k = P\!\cdot k/(\surd{P^2}\surd{k^2})$ and $z_p = P\!\cdot \!p/(\surd{P^2}\surd{p^2})$, and the 
momenta,  $k$ and $p$, are discretized, e.g.,
\begin{equation}
  \mathcal{F}_{P_n}^\alpha (p_i,P) =\sum_{m=0}^\infty \mathcal{F}_{P_n,m}^\alpha (p_i,P) U_m(z_p) \ 
  \label{chebyshev}
\end{equation}
where we define $\mathcal{F}^\alpha_{mi} \equiv \mathcal{F}_{P_n,m}^\alpha (p_i,P)$.  We employ three Chebyshev polynomials
for the ground states and five Chebyshev polynomials for the excited states. 

We solve the eigenvalue problem posed in Eq.~(\ref{eigenvalue}) by means of Arnoldi factorization implemented in the 
\texttt{ARPACK} library~\cite{Arpack,Blank:2010bp} which computes the eigenvalue spectrum for a given $N\times N$ matrix. 
Practical implementation implies a transcription of the BSE kernel $\mathcal{K}^{\alpha\beta} (p,k,P)$ in Eq.~(\ref{KernelAB}) as 
 \begin{align}
   \mathcal{F}^{\alpha}_{m i}= \left (\mathcal{K}^{\alpha \beta} \right )^{mi}_{nj}  \mathcal{F}^{\beta}_{nj} \ ,
  \label{sixindices} 
 \end{align}
which depends on six indices. In order to cast Eq.~(\ref{sixindices}) as an $N\times N$ matrix, 
\begin{align}
  \mathcal{F}_I= \mathcal{K}_{I J} \, \mathcal{F}_J , \quad   I,J = 1,..., N \ ,
\end{align}
we relate the two sets of indices, $I,J$ and $\alpha,\beta,m,n,i,j$, with the condition,
\begin{align}\label{eq:index}
I(\alpha,m,i) & =
    \alpha(m+1)i\ , \\
J(\beta,n,j) & =
   \beta(n+1)j \ .
\end{align}
To recover the components of the BSA, $\mathcal{F}^\alpha_P(p,P)$, one simply identifies $ \mathcal{F}^\alpha_{m i} = \mathcal{F}_I$  
where  $I$ is given by Eq.~(\ref{eq:index}).

As mentioned, the factorization with  \texttt{ARPACK} yields the eigenvalue spectrum,  $\lambda_n (P^2)$, of the kernel, $\mathcal{K}_{IJ}$, 
and therefore we obtain eigenvectors, $\mathcal{F}_I$, or equivalently the BSA, $\mathcal{F}_{P_n}^\alpha(p,P)$, for each
$\lambda_n (P^2=-M_n^2)=1$.\footnote{To find the root  $M_n$ of the equation  $\lambda_n (P^2=-M_n^2)-1=0$, we made use of the 
Numerical Recipe subroutines \texttt{zbrent} and \texttt{rtsec}~\cite{Press:1992zz}. We cross-checked the \texttt{ARPACK} solutions 
for the ground states of the pseudoscalar channel with the commonly used iterative procedure and find excellent 
agreement of the order $10^{-16}$.}  These numerical results are presented and discussed in Sec.~\ref{discussion}.

For the sake of completeness, we mention that all BSA are normalized canonically as
\begin{align}
    2 P_\mu & =  \int_k^{\Lambda}\!  \mathrm{Tr}_\mathrm{CD} \left [   \bar \Gamma_{P_n} (k,-P) \frac{\partial S(k_+)}{\partial P_\mu}
             \Gamma_{P_n} (k,P) S(k_-) \right. \nonumber \\
  &  +  \left.    \bar \Gamma_{P_n} (k,-P) S(k_+) \Gamma_{P_n} (k,P) \frac{\partial S(k_-)}{\partial P_\mu}  \right ]  \ ,
\label{norm}
\end{align}
where we have omitted a third term that stems from the derivative of the kernel, $\partial \mathcal{K}^{kl}_{mn}(p,k,P)/ \partial P_\mu$,
as it does not contribute in the RL truncation of Eq.~(\ref{BSEkernel}).\footnote{We verify the values obtained with Eq.~(\ref{norm})
with the equivalent normalization condition~\cite{Nakanishi:1965zza,Fischer:2009jm}: $(d\ln\lambda_n/dP^2)^{-1}= \mathrm{tr} 
\int_k^\Lambda 3 \bar \Gamma(k,-P) S(k_+)\Gamma(k,P)S(k_-)$.}
In Eq.~(\ref{norm}), the charge-conjugated BSA is defined as $\bar \Gamma (k,-P) := C\, \Gamma^T (-k,-P) C^T$, where $C$ is the charge 
conjugation operator. We shall discuss the peculiarities of charge symmetry in the case of heavy-light mesons in Sec.~\ref{discussion}.

\section{Discussion of results \label{discussion}}

We summarize our results for the mass spectrum and weak decay constants of the pseudoscalar flavor singlet and nonsinglet mesons in 
Tables~\ref{table1}, \ref{table3} and \ref{table2}, where the DSE and BSE are solved for two parameter sets of the interaction in 
Eq.~(\ref{qinchang}), denoted by models 1 and 2 in all tables, also discussed in Ref.~\cite{Qin:2011xq}.  
In Table~\ref{table1}, we list the masses and decay constants, adopting the Particle Data Group (PDG) conventions~\cite{Beringer:1900zz}, 
of the $\pi, K, \pi(1300), K(1460), \eta_c(1S)$ and $\eta_c(2S)$. We also include the mass and decay constant of a pure $\bar ss$ state 
which cannot be related to the $\eta$ or $\eta '$ but whose study is worthwhile to elucidate the trajectory of the weak decay constants 
as a function of the current quark mass. We remind that the $\eta_c$ and $\bar ss$ are neutral particles which do 
not undergo a purely leptonic decay as in Eq.~\eqref{decay}, yet they are useful quantities to calculate~\cite{Davies:2010ip}.

A direct comparison of the mass and decay constant entries in both model columns reveals that the values obtained with model~1 
are in much better agreement with experimental values of the $0^-$ ground states, namely the $\pi$, $K$ and $\eta_c$, whose 
$\omega$ dependence in the range $\omega \in [0.4,0.6]$~GeV is rather weak. Conversely, model~2 reproduces well the masses 
of the radially excited states, $\pi(1300)$, $K(1460)$ and  $\eta_c(2S)$, where for $\omega D = (1.1~\mathrm{GeV})^3$ the ground
states are no longer insensitive to $\omega$ variations~\cite{Qin:2011xq}.  In order to obtain $m_\pi = 0.138$~GeV, 
$\omega$ must increase beyond our reference value, $\omega = 0.6$~GeV, for the excited spectrum.

\begin{table}[t!]
\caption{Mass spectrum and decay constants for flavor singlet and nonsinglet $J^{P}=0^-$ mesons, where we follow 
Particle Data Group conventions~\cite{Beringer:1900zz}. Both models refer to the interaction {\em Ansatz} in Ref.~\cite{Qin:2011dd}, 
where we use the values $\omega = 0.4~\text{GeV}$ and  $\omega D =(0.8~\text{GeV})^3$  for model~1 and $\omega =0.6~\text{GeV}$ 
and $\omega D =(1.1~\text{GeV})^3$ for model~2. Dimensioned quantities are reported in GeV and reference values 
in the last column include experimental averages and lattice-QCD predictions when known. }
\label{table1}
\begin{center}
\begin{tabular}{p{1.7cm}p{2cm} p{2cm}p{2cm}}
\hline\hline
                      & Model 1      &   Model 2       &     Reference     \\
\hline
$m_{\pi}$           &  0.138          &       0.153            &  0.139~\cite{Beringer:1900zz}            \\
$f_{\pi}$             &  0.139          &     0.189            &  0.1304~\cite{Beringer:1900zz}           \\ \hline
$m_{\pi(1300)}$            & 0.990            & 1.414              &  $1.30\pm 0.10$~\cite{Beringer:1900zz}                 \\
$f_{\pi (1300)}$            & $-1.1\times10^{-3}$  & $-8.3\times10^{-4}$  &             \\
$f_{\pi(1300)}\gmor$       &  $-1.4\times10^{-3}$ & $-4.0\times10^{-4}$&                  \\ \hline
$m_K$                 & 0.493            & 0.541             &  0.493~\cite{Beringer:1900zz}           \\
$f_K$                 & 0.164            & 0.214              &  0.156~\cite{Beringer:1900zz}            \\
$f_K\gmor$            & 0.162            & 0.214              &                  \\   \hline
$m_{K(1460)}$          & 1.158            &1.580               &   1.460~\cite{Beringer:1900zz}                \\
$f_{K(1460)}$              & $-0.018$        & $-0.017$             &                  \\
$f_{K(1460)}\gmor$    & $-0.018$       & $-0.017$             &                  \\  \hline
$m_{\bar ss}$       & 1.287            &1.702               &                 \\
$f_{\bar ss}$       &  $-0.0214$      &  $-0.0216$            &                  \\
$f_{\bar ss}\gmor$  & $-0.0215$           & $-0.0218$             &                  \\ \hline
$m_{\eta_c(1S)}$          & 3.065      & 3.210              &  2.984~\cite{Beringer:1900zz}           \\
$f_{\eta_c(1S)}$      & 0.389            & 0.464             &  0.395~\cite{Davies:2010ip}   \\
$f_{\eta_c(1S)}\gmor$ & 0.380      & 0.451              &                  \\  \hline
$m_{\eta_c(2S)}$      & 3.402            & 3.784              &  3.639~\cite{Beringer:1900zz}           \\
$f_{\eta_c(2S)}$      & 0.089           & 0.105              &                  \\
$f_{\eta_c(2S)}\gmor$ & 0.088    & 0.103              &                  \\ 
\hline \hline
\end{tabular}
\end{center}
\end{table}

Nonetheless, the ground states are noticeably less dependent on the $\omega D$ values than the radial excitations where large 
mass differences are observed between both columns. This agrees with the observations made in Ref.~\cite{Qin:2011xq} 
and extends them to the strange and charm sector:  the quantity $r_\omega := 1/\omega$ is a length scale that measures the range 
of the infrared component of the interaction, $\mathcal{G}(q^2)$, in Eq.~(\ref{qinchang}). The radially excited states or exotics 
are expected to be more sensitive to long-range characteristics of $\mathcal{G}(q^2)$ than ground states and we confirm 
that all radially excited states increase in mass if the range, or strength, of the strong piece of the interaction is reduced 
(N.B.: for $\omega =0$ the range is infinite). 

In summary, we do not find a parameter set that describes equally well the entire 
mass spectrum of ground and excited states which hints at the insufficiency of this truncation. However, Model~2 uniformly 
overestimates all masses by $6$\%$-10$~\% in Table~\ref{table1}, where the masses are more ``inflated" for the pion and the kaon. 
It is plausible that radiative and hadronic corrections return them to the observed values~\cite{Eichmann:2008ae}. In the 
mass spectrum, we also find states with unnatural time parity, $J=0^{--}$, referred to as exotics and presented in Table~\ref{table3}. 
The mass difference pattern of these states parallels the ones observed in Table~\ref{table1}. However, as they are not
experimentally observed and expected to have masses above 2~GeV~\cite{Beringer:1900zz}, their appearance in the spectrum
is likely an artifact of the current truncation; see, e.g., discussion in Ref.~\cite{Fischer:2014xha}. 

\begin{table}
\caption{List of exotic states with ``unnatural time parity", $J=0^{--}$. In case of the $\bar us$ meson, $C=-1$, must be understood
as approximate, since $|\bar u s \rangle$ is not an eigenstate of charge conjugation. Models~1 and 2 are as in Table~\ref{table1}
and dimensioned quantities are in GeV.  The decay constants of flavorless states are of the order $10^{-6}$~GeV.}
\begin{center}
\label{table3}
\def\arraystretch{1.5}
\begin{tabular}{p{1.8cm}p{1.5cm}p{1.5cm}c}
\hline\hline
                                         & Model 1    &  Model 2        \\   \hline
$m_{\bar uu} $    &   0.733       &  1.049           \\
$m_{\bar ss}$      &   0.856      &   1.351          \\ 
$m_{\bar cc}$      &   3.243      &   3.515          \\ 
$m_{\bar u s}$     &   0.917      &   1.225          \\
$f_{\bar us}$        &  0.0152    &    0.0098         \\
$f_{\bar us}\gmor$   &    0.0150   & 0.0097        \\
\hline \hline
\end{tabular}
\end{center}
\vspace*{-3mm}
\end{table}

\begin{figure}[t!] 
\centering
  \vspace*{-6mm}
   \includegraphics[scale=0.34]{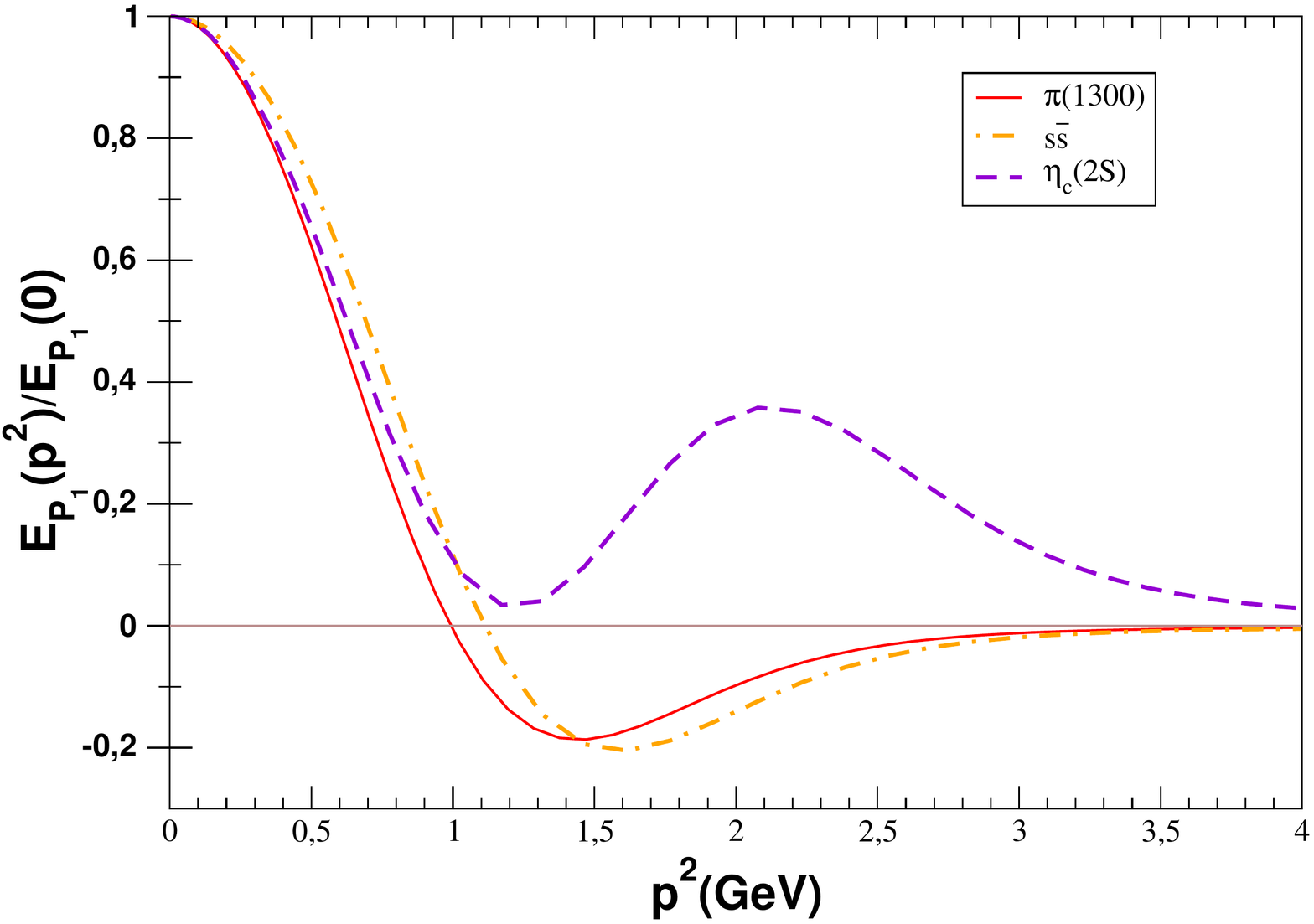} 
   \caption{Lowest Chebyshev moment, $^0E_{P_1}(p^2)$, associated with the leading Dirac structure $E_{P_1}(p^2)$
   of the meson's BSA~\eqref{diracbase} for the first radial excitations $\pi(1300)$, $\eta(1760)$ and $\eta_c(2S)$. The interaction
   parameters are $\omega = 0.6$ and $\omega D = (1.1~\mathrm{GeV})^3$.     }
   \label{Chebyshevfigure}
   \vspace*{-3mm}
\end{figure}

The weak decay constants are obtained via Eq.~(\ref{decay}) and verified by means of the Gell-Mann--Oakes--Renner (GMOR) 
relation~\eqref{eq:gmor},
a relation valid for every $0^-$ meson irrespective of the magnitude of the current-quark mass, $m^{f,g}(\mu)$. In the chiral limit, 
\begin{equation}
  \rho _{P_n}^0 (\mu) :=   \lim_{{\hat m} \to 0} \, \rho_{P_n} (\mu) < \infty \ ,  \  \forall n , 
\end{equation}
owing to the ultraviolet behavior of the quark-antiquark scattering kernel in QCD which guarantees that $\rho_{P_n} (\mu )$ in Eq.~(\ref{rho}) is cutoff
independent. A necessary corollary is that in the chiral limit the decay constant of excited states vanishes identically~\cite{Holl:2004fr}:
\begin{equation}
  f _{P_n}^0 (\mu) \equiv  0  \ ,  \  n \geq  1  \ .
\label{decayexcited}
\end{equation}
We provide numerical verification of Eq.~(\ref{decayexcited}) in the case of the first radial excitation of the pion, i.e. the parity 
partner $\pi (1300)$, in Table~\ref{table1} from which it can also be read that the decay constant of the parity partner of the kaon,  
the $K(1460)$, is strongly suppressed. Analogously, the decay constant value of the $\bar ss$ state is 
very small. On the other hand, we note that all weak decay constants of the excited states but the $\eta_c(2s)$ are 
negative with model~1, slightly less so when using model~2, which  is consistent with the discussion in Ref.~\cite{Qin:2011xq}.
As it becomes clear from Fig.~\ref{Chebyshevfigure}, for $\omega \geq 0.5$ and  $p^2 \gtrsim 1$~GeV$^2$, the leading amplitude's
lowest Chebyshev projection, $^0 E_{P_1}(p^2)$,  is negative definite in case of the first radial excitations,  $\pi(1300)$ and $\bar ss$, 
whereas it remains positive for the ground states, which parallels the pattern of wave functions in quantum mechanics.  
A necessary consequence is then  $f_\pi >0$ and $f_{\pi(1300)} < 0$, also observed for the kaon and its first radial excitation 
and for the $\eta(1760)$. The negative value for $f_{\pi(1300)} $ is consistent with lattice-QCD simulations~\cite{McNeile:2006qy}. 
The equality in Eq.~(\ref{decayexcited}) is only valid in the chiral limit and for increasing current-quark masses the excited pseudoscalar 
meson's decay constant first remains negative, but after a minimum, which occurs between the strange and charm-quark mass,  
steadily increases and becomes positive before reaching the charm quark mass.  We therefore find for the  $\eta_c(2S)$ a positive 
decay constant for both model parameters. We note that our turning point is  below that observed in  Ref.~\cite{Krassnigg:2006ps},
which is about the charm-quark mass. 

We now turn our attention to the charmed pseudoscalars, in particular the $D$ and $D_s$ mesons. Before discussing the results, 
a few technical comments are in order. Besides being bounded from above, the eigenvalue spectrum of the BSE should be positive 
definite~\cite{Ahlig:1998qf} owing  to the Hermiticity requirement of physical operators. However, this is not generally true for flavored
mesons with unequal masses, in particular heavy-light systems in the rainbow-ladder truncation.  For example, in solving the BSE 
for the $D$ and $D_s$ mesons, we find a complex conjugate pair of eigenvalues for the excited $D_{(s)}$ states if we include contributions 
from higher Chebyshev polynomials, $U_{m>1} (z_p)$. This occurs for either parameter set of the interaction and indicates a
non-Hermiticity of the Hamiltonian in the case of $|\bar cu\rangle$ and $|\bar cs\rangle$ bound states.\footnote{A discussion of complex 
eigenvalues and their origin in the crossing of normal and abnormal (which have vanishing binding energy) eigenstates can be 
found in Ref.~\cite{Ahlig:1998qf}. Note that in quantum mechanics even non-Hermitian operators yield real and positive eigenvalues 
provided that $PT$ symmetry is conserved, whereas if it is broken the eigenvalue spectrum is complex~\cite{Bender:2007nj}. 
Yet, an analogical observation in Quantum Field Theory and more precisely in nonperturbative QCD is not straightforward.} 
We have verified that our solutions converge, i.e. they are independent of the amount of Chebyshev moments in the
expansion and the ground state and excited $D$ meson masses do not vary for $U_m (z_p), m \geq 4$.

We remind that nonequal mass mesons, such as the $K$ and $D$, are not eigenstates of the  charge-conjugation operator 
and thus $\bar \Gamma (k,P) = \lambda_c \Gamma (k,P)$ does not imply $\lambda_c =\pm 1$ for the charge parity. On the other 
hand, for  equal-mass ($\bar uu, \bar dd, \bar ss, \bar cc$) pseudoscalar mesons with $J^{PC} = 0^{-+}$, the  constraint that  the 
Dirac base  satisfies $\lambda_c = + 1$ requires the dressing functions, $\mathcal{F}_{P_n}^\alpha (p,P)$,  to be {\em even\/} 
in  the angular variable $z_p$. In the case of the $D$ and $D_s$ mesons, however, we do observe that also {\em odd\/}
Chebyshev moments contribute and they acquire an imaginary part in the ground and excited states. The ground-state 
eigenvalues of the $D$ mesons remain real and though the associated eigenstate is a solution of the BSE, it cannot 
be interpreted as a physical solution for the $D$ mesons due to the non-Hermiticty invoked above. If we limit the Chebyshev expansion 
to the lowest order, the BSA is independent of the angle, $z_p$, and therefore remains real in all cases as are the eigenvalues.  
Hence, it is the angular-dependent higher-order terms of the Chebyshev expansion that lead to complex eigenvalues for the 
charmed excited states within the RL truncation. On the other hand, this does not occur for the kaon where flavor-symmetry and 
charge-parity breaking are still negligible.

\begin{table}
\caption{Charmed meson observables computed in the lowest-order Chebyshev moment approximation (see text for explanations). 
Models~1 and 2 are as in Table~\ref{table1}. Dimensioned quantities are reported in GeV and experimental values are 
averages from the Particle Data Group. }
\begin{center}
\label{table2}
\begin{tabular}{lccc}
\hline \hline
                      &  Model 1      &  Model 2      &     Experiment~\cite{Beringer:1900zz}      \\  \hline
$m_{D}$         &  2.115       & 2.255  &   1.869              \\ 
$f_{D}$           &  0.204       & 0.281  &   $ 0.2067 \pm 0.0085 \pm 0.0025$                \\                                          
$m_{D_s}$     &  2.130       & 2.284  &  1.968           \\ 
$f_{D_s}$       &  0.249       & 0.320  &  $0.260 \pm 0.005$           \\
\hline\hline
\end{tabular}
\end{center}
\end{table}

The mass and decay constant entries in Table~\ref{table2} are all obtained in the lowest-order Chebyshev approximation which is 
independent of the angle, $z_p$. Incidentally, this approximation bears similarity with common phenomenological {\em Ans\"atze\/} to the 
BSA for $D$ mesons~\cite{Ivanov:1997yg,Ivanov:2007cw,ElBennich:2010ha,ElBennich:2011py,ElBennich:2012tp} and this is exactly
what the values in Table~\ref{table2} represent: masses and decay constants obtained with a model based on the lowest-order 
approximation for which the eigenvalues and BSA are real. The mass difference, $\Delta m \simeq 15$--$30$~MeV, between the $D$ and 
$D_s$ mesons is smaller than experimentally observed, i.e. $\Delta m \sim 100$~MeV. On the other hand, the weak decay constants 
are in good agreement with experimental averages for model~1 which again provides the preferred interaction for the ground states. 
We also stress that the results in Table~\ref{table2} were obtained without any modification of the interaction~\eqref{qinchang} in the 
RL truncation, such as suppression of the infrared domain~\cite{Nguyen:2010yj} or use of a constituent-quark mass~\cite{Souchlas:2010zz}. 
Nevertheless, as has been realized previously~\cite{Nguyen:2010yj}, the RL approximation is an inadequate truncation scheme for heavy-light mesons.

\section{Conclusion}

We computed the BSAs for the ground and first excited states of the flavor singlet and flavored pseudoscalar mesons with an interaction
{\em Ansatz} that is massive and finite in the infrared and massless in the ultraviolet domain. This interaction is qualitatively in accordance 
with the so-called {\em decoupling solutions\/} of the gluon's dressing function and thus represents an improvement on the Maris-Tandy 
model~\cite{Maris:1999nt}. In conjunction with the RL truncation, the latter proves to be a successful interaction model for the light meson 
spectrum, $M\lesssim 1$~GeV, but fares less well in applications to heavy-light systems where infrared modifications of the interaction
are required in order to obtain numerical results which are still not satisfactory~\cite{Nguyen:2010yj}. 

Motivated by the successful application of the interaction {\em Ansatz} of Eq.~\eqref{qinchang} to the mass spectrum of light mesons as well as 
some of their excited states in Ref.~\cite{Qin:2011xq}, we extend this study to the strange and charm sectors and obtain the masses of ground 
states and resonances as well as decay constants as presented in Table~\ref{table1}. These numerical results are in good agreement
with experimental averages for the ground states, yet we confirm the earlier observation that no single parametrization of Eq~\eqref{qinchang}
is able to reproduce the mass spectrum of both, the ground and excited states in the RL truncation. The case of the $D$ and $D_s$ mesons 
remains equally unresolved and clearly shows that the mass asymmetry and disparate scales in heavy-light systems require corrections beyond 
the leading truncation. We postpone the treatment of heavy-flavored mesons with a BSE {\em Ansatz} valid for any symmetry-preserving dressed 
quark-gluon vertex~\cite{Chang:2009zb} to a future publication. We stress that the merit of BSE solutions for heavy-light systems goes beyond 
a successful description of the heavy meson's mass spectrum and decay constants and that the light-front projection of the meson's Bethe-Salpeter 
wave function allows to extract its light cone distribution amplitudes~\cite{Chang:2013pq,Shi:2014uwa}. The latter, commonly expanded in 
Gegenbauer polynomials, encode the relevant nonperturbative information in QCD factorization of heavy-meson weak decays and their calculation 
represents a big step beyond the usually employed asymptotic form, $\phi (x) \propto 6x(1-x)$. Indeed, in the case of $D$ and $B$ mesons, very little 
is known about the nonperturbative nature of these distribution amplitudes.

\acknowledgements 
This work was supported by the S\~ao Paulo Research Foundation (FAPESP) and benefitted from access to the computing facility of the 
Centro Nacional de Supercomputa\c{c}\~ao at the Federal University of Rio Grande do Sul (UFRGS). We appreciated insightful discussions 
with and comments by Lei Chang, Gernot Eichmann, Andreas Krassnigg, Craig Roberts and Peter Tandy.

\appendix

\section{Renormalization conditions  \label{renorm}}

In order to eliminate the cutoff dependence, it is necessary  to impose renormalization conditions. The usual procedure 
is to relate bare and renormalized masses as in Eq.~(\ref{mzeta}) and for the fermion fields one writes analogously,
\begin{equation}
  q_f^{\text{bm}}(p^2) = \left ( Z_2^f(\mu,\Lambda) \right )^{\!\frac{1}{2}} \! q_f(p^2,\mu) \ . 
\end{equation}
The renormalization constants $Z_m$ and $Z_2$ are determined by imposing the  conditions, 
\begin{equation}
\label{eq:rc}
   B(\mu,\mu) = m_f(\mu) \ \ \mathrm{and} \ \ A(\mu,\mu) =  1.
\end{equation}
For arbitrary flavors and using Eqs.~(\ref{mzeta}) and (\ref{eq:rc}), the GMOR expression can be generally written as, 
\begin{align}
\label{eq:gmor}
     f_{P_n}M_{P_n}^2 & =  \left (m_f^{\text{bm}}+m_g^{\text{bm}} \right )\rho_{P_n}^{\text{bm}}  \nonumber \\
                  =  &  - \left( m_f^\text{bm}+m_g^{\text{bm}}\right )\,  \langle0| \bar{q}_g^{\text{bm}}\gamma_5 q_f^{\text{bm}}|P_n \rangle
         \nonumber   \\
                  = & - \left(Z_m^f m_{g}(\mu)+Z_m^g m_{f}(\mu)\right)  (Z_2^f Z_2^g)^{1/2}
                           \langle0|\bar{q}_g\gamma_5 q_f|P_n \rangle
         \nonumber \\
                  = &\left( \sqrt{\frac{Z_m^f}{Z_m^g}}\,  m_{g}(\mu)+\sqrt{\frac{Z_m^g}{Z_m^f}}\,  m_{f}(\mu)\right)
                          \rho_{P_n} (\mu)    \ ,
\end{align}
where we have,
\begin{align} 
\label{rho}
       \rho_{P_n} (\mu )  &=  - (Z_4^f Z_4^g)^{\frac{1}{2}}  \langle 0 |\, \bar{q}_g\gamma_5 q_f |  P_n \rangle   \\
        =  &  -i\, (Z_4^f Z_4^g)^{\frac{1}{2}}\, \text{Tr}_\mathrm{CD}\! \int_k^\Lambda\! \gamma_5\,  S_f (k_+)\Gamma_{P_n}^{fg}(k,P) S_g (k_-) \ ,
        \nonumber
\end{align}
with $Z_4^f(\mu,\Lambda)=Z_2^f(\mu,\Lambda)Z_m^f(\mu,\Lambda)$. The weak decay constant can, of course, be directly inferred from, 
\begin{align}
 \label{decay}
\hspace*{-2mm}
   f_{P_n}P_{\mu}  &= -i \, \left  \langle 0 \left  | \bar{q}_g^{\text{bm}} \gamma_5 \gamma_\mu q_f^{\text{bm}} \right |\! P_n \right \rangle  \\
  = & \ (Z_2^f Z_2^g)^{\frac{1}{2}}\, \text{Tr}_{\text{CD}}\! \int_k^\Lambda \! \gamma_5 \gamma_\mu\,  S_f (k_+) \Gamma_{P_n}^{fg} (k,P)S_g (k_-) \, .
  \nonumber 
\end{align}

In solving the DSE within a complex parabola as discussed in Sec.~\ref{complexplane}, it is convenient to compute the renormalization 
constants at a momentum scale, $\xi$, different from $\mu > \xi$ while still demanding 
that Eq.~\eqref{eq:rc} be satisfied. Once we know the solutions of and $A (p^2)$ and $B(p^2)$ renormalized at $p^2 = \mu^2$, the values  
for $B(\xi,\mu)$  and $A(\xi,\mu)$ are readily inferred. One may as well calculate $Z_m^f(\mu,\Lambda)$ and $Z_2^f(\mu,\Lambda)$ 
at a point, $p^2=\xi^2$, using the condition,
\begin{align}
\label{eq:rc2}
 B(\xi,\mu) & = Z_4(\mu,\Lambda)\,  m_f(\mu)+\Pi_B(\xi,\Lambda) \ ,  \notag\\
 A(\xi,\mu)  &  = Z_2(\mu,\Lambda)+\Pi_A(\xi,\Lambda) \ ,
\end{align}
where $\Pi_B$ and  $\Pi_A$ are the expressions inferred from the appropriate projections of the quark self-energy
correction in Eq.~(\ref{dressedquark}). Solving the DSE starting by imposing new renormalization conditions, namely
$B(p^2,\xi)$  and $A(p^2,\xi)$,  in Eq.~(\ref{eq:rc2}), one verifies that the conditions in Eq.~(\ref{eq:rc}) are
satisfied with high precision.
We check the stability of all results from $\Lambda =4$~GeV up to $\Lambda=10$~GeV, in which case $Z_2(\xi,\Lambda)$ and 
$Z_4(\xi,\Lambda)$ are modified in order to absorb the cutoff dependence; our results prove to be stable and show no cutoff 
dependence. The GMOR relation (\ref{eq:gmor}) provides us with an additional cross-check.





\end{document}